# Quantum Behavior of Water Molecules Confined to Nanocavities in Gemstone

*Boris P. Gorshunov*[1,2,3,*], *Elena S. Zhukova*[1,2,3], *Victor I. Torgashev*[4], *Vladimir V. Lebedev*[3,5], *Gil'man S. Shakurov*[6], *Reinhard K. Kremer*[7], *Efim V. Pestrjakov*[8], *Victor G. Thomas*[9], *Dimitry A. Fursenko*[9], *and Martin Dressel*[2,*]

1 A.M. Prokhorov General Physics Institute, Russian Academy of Sciences, Vavilov str. 38, 119991 Moscow, Russia

2 1. Physikalisches Institut, Universität Stuttgart, Pfaffenwaldring 57, 70550 Stuttgart, Germany

3 Moscow Institute of Physics and Technologyy, 141700, Dolgoprudny, Moscow Region, Russia

4 Faculty of Physics, Southern Federal University, 344090 Rostov-on-Don, Russia

5 Landau Institute for Theoretical Physics, Russian Academy of Sciences, Chernogolovka, Moscow Region, Russia

6 Kazan Physical-Technical Institute, Russian Academy of Sciences, 10/7 Sibirsky trakt, 420029 Kazan, Russia

7 Max-Planck-Institut für Festkörperforschung, Heisenbergstraße 1, 70569 Stuttgart, Germany

8 Institute of Laser Physics, Russian Academy of Sciences, 13/3 Ac. Lavrentyev's Prosp., 630090 Novosibirsk, Russia

9 Institute of Geology and Mineralogy, Russian Academy of Sciences, 3 Koptyug st., 630090, Novosibirsk, Russia

**Corresponding Authors**

*Boris Gorshunov: gorshunov@ran.gpi.ru*
*Martin Dressel: dressel@pi1.physik.uni-stuttgart.de*

SUBJECT CATEGORIES

Spectroscopy

Surfaces, porous materials

ABSTRACT

When water is confined to nanocavities, its quantum-mechanical behavior can be revealed by terahertz spectroscopy. We place $H_2O$ molecules in the nanopores of a beryl crystal-lattice and observe a rich and highly anisotropic set of absorption lines in the terahertz spectral range. Two bands can be identified, which originate from translational and librational motions of the water molecule isolated within the cage; they correspond to the analogous broad bands in liquid water and ice. In the present case of well-defined and highly symmetric nanocavities, the observed fine structure can be explained by macroscopic tunneling of the $H_2O$ molecules within a six-fold potential caused by the interaction of the molecule with the cavity walls.

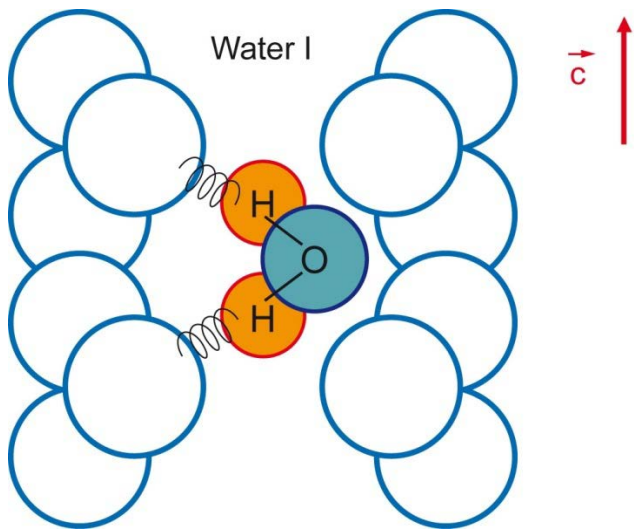

Water is the chemical compound most pervasive on Earth, the liquid most studied in physics, chemistry and biology.[1] Nevertheless, despite the simple structure of a $H_2O$ molecule, there is still a lot to be understood because of its complex network of hydrogen bonds. Once in contact with solid surfaces[2,3] or single macromolecules, for instance, proteins,[4,5] water forms a hydration layer in which a gradual transition from bulk towards bound water takes place. In the course of this crossover, the characteristic network becomes less dynamic, but the weak hydrogen bonds still allow librations, i.e. restricted rotations of the water molecules, which survive even in solid ice. Here we go one step further to investigate single water molecules confined to nanoscale cavities[6] present in the crystalline gemstone, called beryl. The mineral beryl (the chemical formula is $Be_3Al_2Si_6O_{18}$) is found rather frequently in various deposits around the world and forms a large family of gemstones (emerald, aquamarine, heliodor, morganite, etc.). The difference in colors is determined by replacing Al or Be with other cations (Mn, Mh, Fe). Beryl belongs to the cyclo-silicates, which connect $SiO_4$ tetrahedra as building blocks to $Si_6O_{18}$ rings, as shown in Figure 1. Staggered along the crystallographic *c*-axis, they form chains of nanocavities that can host single water molecules. Due to their distance from each other they can be considered to be isolated. The dipole moment of these $H_2O$ molecules is either perpendicular (called type-I water) or parallel (type-II water) to the *c*-axis, as sketched in Figrue1b, c. Our optical measurements reveal quantum behavior of the type-I molecules that are loosely bound to the walls of the beryl nanopore.

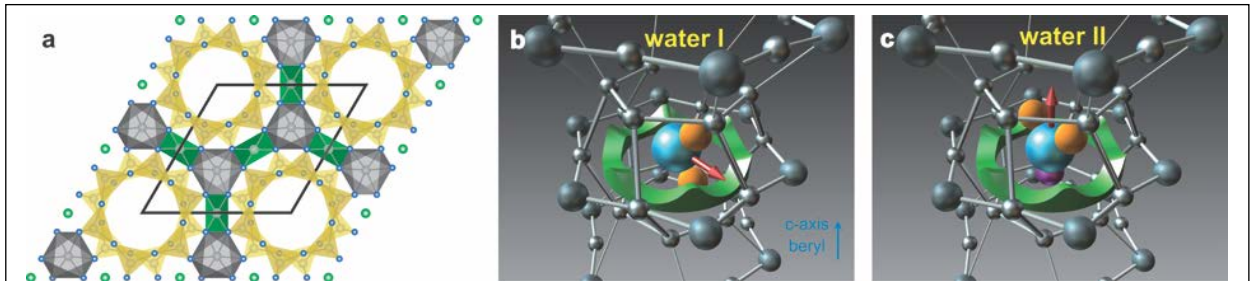

**Figure 1. (a)** Honeycomb crystal structure (space group P6/mcc) of beryl in the plane perpendicular to the *c*-axis. Stacked six-membered rings of $SiO_4$ form large open nanochannels that extend along the *c*-direction. They contain cavities of 5.1 Å diameter connected by bottlenecks of 2.8 Å which can be clogged by alkali ions (Na or K). Crystal water molecules within the cavities are oriented with the dipole moment either perpendicular (**b**) or parallel (**c**) to the *c*-direction. The type-II molecules are rotated by 90° relative to the molecules of type-I due to the Coulomb interaction with the positively charged alkali ions. Hydrogen bonds weakly connect the water molecules to surrounding oxygen atoms of the $SiO_4$ cage[7,8] as indicated by the green belt that resembles the periodic potential. Compared to the behavior of free $H_2O$ molecules, these weak bonds lead to modifications of the dynamical properties. The qualitatively new dynamics are prototypical for the hydrogen-bonded network in solid ice or even liquid water. The red arrows indicate the dipole moments of the $H_2O$ molecules, which are subject to librations and translations. The coupling between the electric field $E$ and the dipole moment is maximal when the two vectors are perpendicular to each other; no coupling takes place for parallel orientation. Light polarized parallel to *c* couples to type-I molecule, whereas light $E \perp c$ probes translations of both types of molecules.

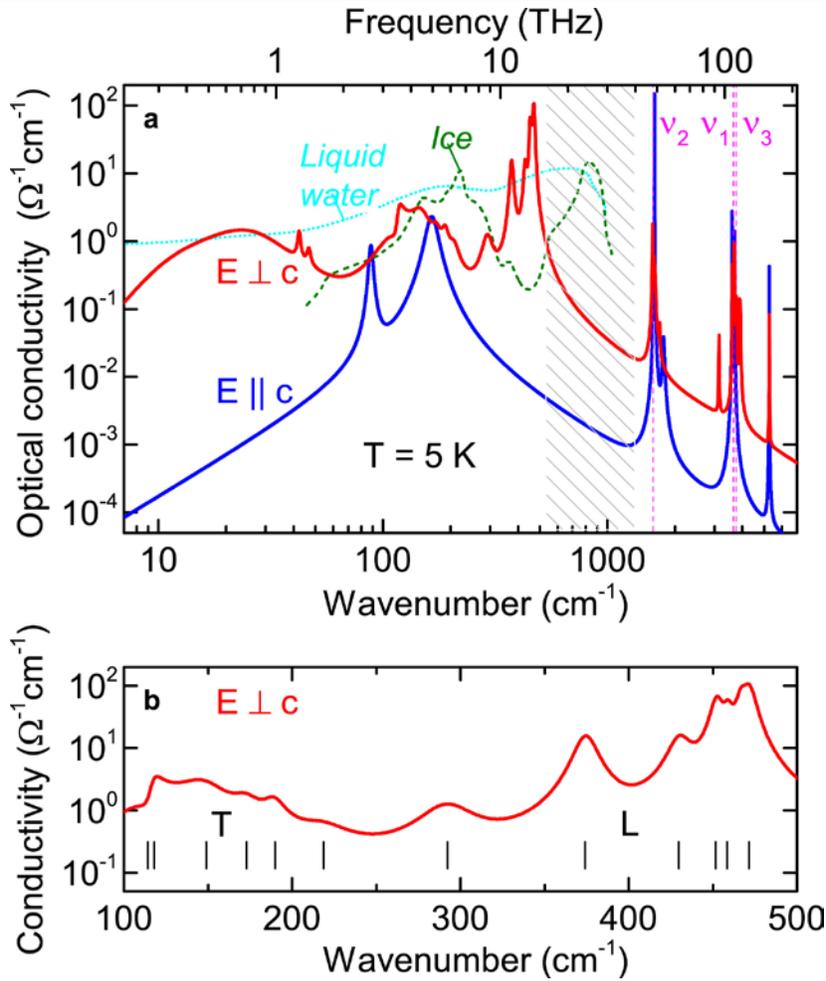

**Figure 2.** (a) Double-logarithmic representation of the optical response of water molecules in beryl nanopores. The optical conductivity spectra are measured for the electric field polarized parallel and perpendicular to the *c*-axis at $T = 5$ K. The hatched area from 500 cm$^{-1}$ to 1200 cm$^{-1}$ is dominated by strong phonon absorption of the beryl lattice, which we subtracted in order to focus on the water-related features (see Supporting information). The magenta lines correspond to the three intra-molecular modes of the H$_2$O molecule. For comparison, spectra of liquid water ($T = 27$°C, dotted cyan line) and hexagonal ice ($T = 100$ K, dashed green line) are included.[9-12] (b) The far-infrared spectral range displayed on a linear scale. The transitions from the ground band to the translational band (T) and to the librational band (L) are indicated.

We have conducted optical transmission and reflection measurements of synthetic beryl single crystals doped with Mn (synthetic morganite) over a wide frequency range from sub-terahertz via terahertz to the infrared band employing several spectrometers, as described in more detail in the Supporting Information. The experiments were performed at different temperatures from $T = 300$ K down to 5 K using light polarized parallel and perpendicular to the nanocavity chains. Reference spectra collected on dehydrated samples enables us to exclude the phonon absorption of beryl and other contributions[13-15] and to unambiguously identify features related exclusively to water (see Supporting Information). The low-temperature spectra of the optical conductivity (that is proportional to the absorption) of the water-related absorption are presented in Fig. 2 for the polarizations $E \parallel c$ and $E \perp c$. For these principal directions, the response is extremely anisotropic, especially below 100 cm$^{-1}$ where the spectra differ by more than three orders of magnitude. This pronounced anisotropy comes rather unexpectedly since we solely consider water-related absorption features. At higher frequencies, $\nu > 1000$ cm$^{-1}$, both spectra ($E \parallel c$ and $E \perp c$) are dominated by three well-known absorption features related to internal H$_2$O modes shown in Fig. 3. The observed peaks $\nu_i$ ($i = 1,2,3$) are slightly shifted relative to those of the free H$_2$O molecule, which have been detected at $\nu_1 = 3656.65$ cm$^{-1}$, $\nu_2 = 1594.59$ cm$^{-1}$ and $\nu_3 = 3755.79$ cm$^{-1}$.[16] As seen in Fig. 3a-c, all three vibrations are excited for either polarization; we conclude that both type-I and type-II water molecules are present in the nanopores. For both spectra, $E \parallel c$ and $E \perp c$, a peak at approximately 5300 cm$^{-1}$ is found and ascribed to the combined vibration $\nu_1 + \nu_2$. Most important are the numerous side bands observed around the internal vibrations; they are combinations of $\nu_i$ and lower frequency resonances.

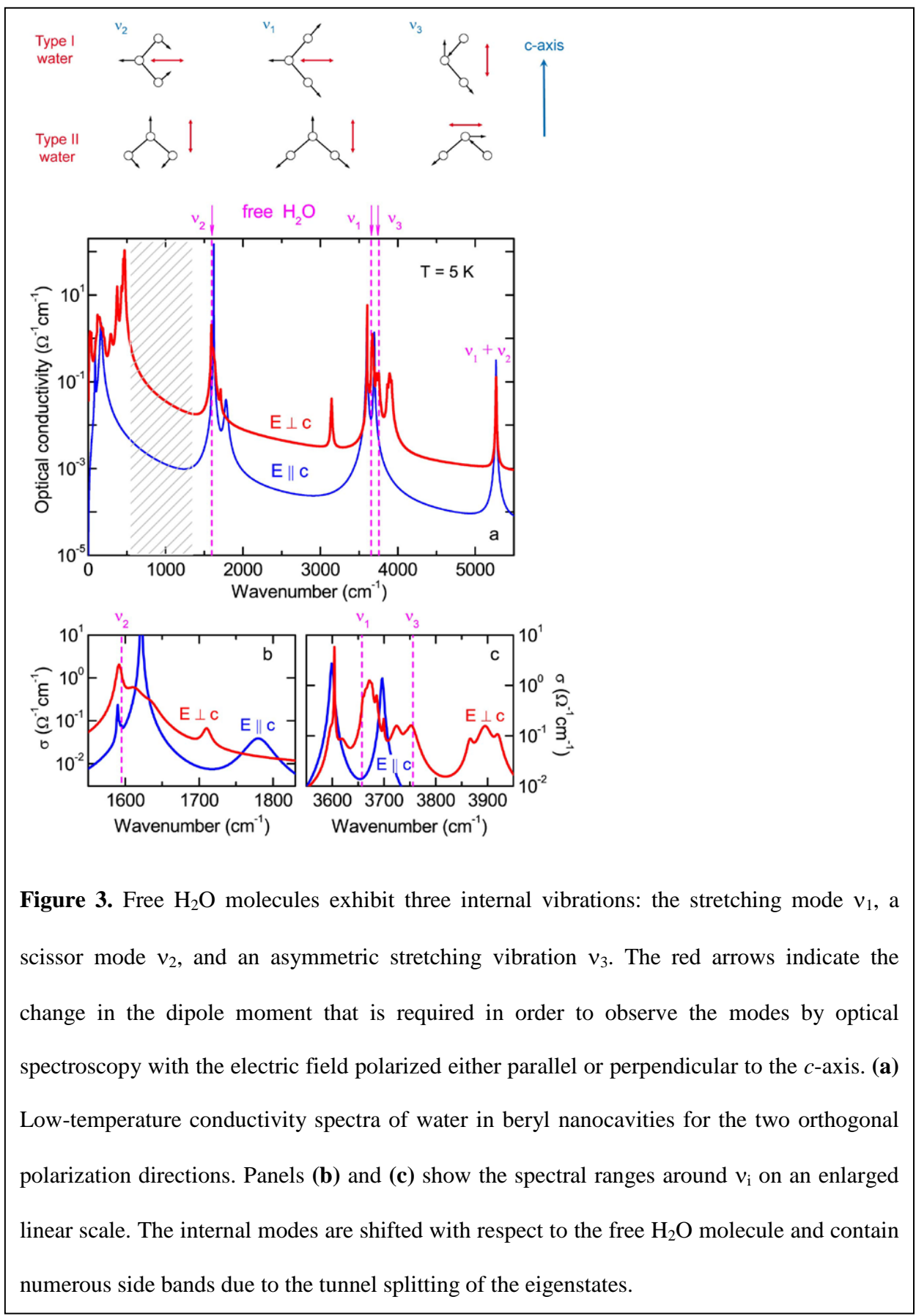

**Figure 3.** Free $H_2O$ molecules exhibit three internal vibrations: the stretching mode $\nu_1$, a scissor mode $\nu_2$, and an asymmetric stretching vibration $\nu_3$. The red arrows indicate the change in the dipole moment that is required in order to observe the modes by optical spectroscopy with the electric field polarized either parallel or perpendicular to the *c*-axis. **(a)** Low-temperature conductivity spectra of water in beryl nanocavities for the two orthogonal polarization directions. Panels **(b)** and **(c)** show the spectral ranges around $\nu_i$ on an enlarged linear scale. The internal modes are shifted with respect to the free $H_2O$ molecule and contain numerous side bands due to the tunnel splitting of the eigenstates.

The focus of the present study, however, is on water vibrations with much lower vibrational energies. In liquid water or in ice, the low-frequency (terahertz to far-infrared) absorption originates from movements of water molecules that are interconnected via a complex network of hydrogen bonds. The corresponding broad absorptions are seen around 200 cm$^{-1}$ (translation, or T-band) and 700 cm$^{-1}$ (libration, or L-band),[9-12] reproduced in the spectra of Figure 2. In the case of H$_2$O confined to highly symmetrical (crystalline) nanopores, the geometry of the H-bonds, that couple the molecule to the walls of the cavity,[8,17] is well defined and thus allows for a detailed description and better understanding of the dynamical properties of the H$_2$O molecules. As seen in Figs. 2 and 3, the optical response of H$_2$O in beryl is very different for the two orientations: the spectrum collected with *E* parallel to the *c*-direction contains only two absorption bands at 88 cm$^{-1}$ and 158 cm$^{-1}$. This can be attributed to the response of type-I water molecules with a dipole moment perpendicular to the *E*-component of the radiation as required for a finite coupling. The type-II water molecules, on the other hand, are not probed since their dipole moments point parallel to the components of the electric field. The corresponding vibrations can be called *c*-axis translations, that is, a shift of type-I H$_2$O along the *c*-axis, and *c*-axis librations, that is to say, a turn of the dipole moment that generates oscillating components on the *c*-axis. In analogy to liquid water or ice, we assign the lower (88 cm$^{-1}$), and the higher (158 cm$^{-1}$) frequency resonances to the *c*-axis translation and *c*-axis librations of type-I H$_2$O, respectively. A much richer spectral structure is observed for the perpendicular polarization *E* ⊥ *c*, as displayed in Fig. 2. A broad maximum is detected at 25 cm$^{-1}$ with two narrower resonance lines at its high-frequency shoulder (cf. Fig. 5). In addition, there are two broad bands above 100 cm$^{-1}$ centered at approximately 150 and 400 cm$^{-1}$, each composed of several narrower features ν$_j$.

Table 1: Parameters obtained from the Lorentz fit of the water related terahertz and far-infrared modes in beryl observed at T = 5 K for the electric field polarized parallel and perpendicular to the c-axis. $\nu_0$ denotes the center frequency, $f$ the oscillator strength, and $\gamma$ the damping. For more details see Supporting Information.

| $\nu_0$ (cm$^{-1}$) | $f$ (cm$^{-1}$) | $\gamma$ (cm$^{-1}$) | Assignment (n,m) → (n,m) |
|---|---|---|---|
| Polarization E ∥ c | | | |
| 88 | 200 | 4 | Water I translational mode |
| 158 | 2 780 | 22 | Water I librational mode |
| Polarization E ⊥ c: THz band (Water I) | | | |
| 10.7 | 106 | 12 | (1,0) → (1,1) |
| 25.9 | 1770 | 25 | (1,1) → (1,2) |
| 42 | 74 | 1.5 | ? |
| 47 | 134 | 6 | ? |
| Polarization E ⊥ c: Far-infrared translational band (Water I) | | | |
| 113 | 3 320 | 43 | (1,1) → (2,0) |
| 117 | 2 960 | 8 | (1,2) → (2,1) |
| 148 | 1 860 | 34 | (1,0) → (2,1) |
| 172 | 900 | 19 | (1,3) → (2,2) |
| 189 | 780 | 15 | (1,1) → (2,2) |
| 218 | 1 300 | 42 | (1,2) → (2,3) |
| Polarization E ⊥ c: Far-infrared librational band (Water I) | | | |
| 292 | 2 020 | 34 | (1,1) → (2,0) |
| 374 | 13 220 | 15 | (1,2) → (2,1) |
| 430 | 10 750 | 14 | (1,0) → (2,1) |
| 452 | 26 600 | 8 | (1,3) → (2,2) |
| 459 | 10 110 | 5.9 | (1,1) → (2,2) |
| 472 | 42 870 | 9 | (1,2) → (2,3) |

Table 1 contains the eigenfrequencies and other parameters of the resonances obtained from a Lorentz fit of the THz and far-infrared spectra for both polarizations, E ∥ c and E ⊥ c. In Fig. 3, we see that many of these frequencies $\nu_j$ show up again as sidebands of the three internal vibrations $\nu_i \pm k\nu_j$ (with $k = 1, 2, \ldots$), i.e. absorption peaks related to combinations of internal high-frequency $\nu_i$ and external low-frequency vibrations $\nu_j$ of $H_2O$.[8] The Supporting Information contains more detailed tables of the numerical values of eigenfrequency, oscillator strength and damping for all the resonances and their assignments.

The rich set of absorption lines in the $E \perp c$ spectra can be consistently explained by a weak, but noticeable coupling of the type-I $H_2O$ molecule to the host crystal-lattice via hydrogen bonds.[8,17] It is reasonable to assume that this coupling of a hydrogen atom to an oxygen atom of the $SiO_4$ cage is not much different in strength from the coupling via H-bonds between $H_2O$ molecules in liquid water or in ice. In view of the overall similarity of the T and L bands observed in liquid water and especially in ice to the two infrared absorption bands found in beryl (Fig.2), we propose that these two bands, around 150 and 400 cm$^{-1}$, are associated with translational and librational motions of the type-I water molecules within the nanocavities of the beryl crystal. These movements can be labeled in-plane translation, that is, the $H_2O$ molecule shifts back and forth as a whole within the plane perpendicular to the *c*-axis, and in-plane libration, where the dipole moment spins around *c*-axis within the plane.

The fact that we have confined single $H_2O$ molecules to the well-defined environment of the nanocavities in a crystal now helps us to understand the vibrational states of water by considering a simple model. According to the six-fold symmetry of the nanopore, the six $H_2O$ equilibrium positions are energetically equivalent. At low temperatures, when thermal excitations become unlikely, the molecule can still tunnel through the barrier separating the energy minima. The Hamiltonian corresponding to in-plane librations of $H_2O$ can be written as

$$H = \frac{\hbar^2}{2I}\frac{\partial^2}{\partial \phi^2} + U(\phi)$$

where $I$ is the moment of inertia of the water molecule and $U$ is the potential that depends on the rotation angle $\phi$ observing a $2\pi/6$ symmetry: $U(\phi) = U(\phi + \pi/3)$. We can classify the quantum mechanical states of type-I water molecules according to how the wave function $\psi$ behaves relative to the $c$-axis rotation by the angle $\pi/3$. When transformed, $\psi$ acquires some extra factor $\exp\{i\phi\}$ with the phase $\phi = m\pi/3$ ($m = 0, \pm 1, \pm 2, 3$). In other words, all states of molecular motion are characterized by their angular number $m$ and can be interpreted as superpositions of the quantum states corresponding to small librations of the type-I water dipole moment near one of the six preferred directions. Corresponding arguments can be applied to the translational states.

Each in-plane librational and similarly translational state is split into six energy levels, among which two of them are twice degenerate, due to mirror symmetry of the potential, as depicted in Fig. 4. The transitions from the ground band to the first excited translational state (T-band) and to the first excited librational state (L-band) can be induced by light according to the selection rule $\Delta m = \pm 1$ imposed by symmetry. This leads to six allowed transitions in each band, from the ground band to the T-band and from the ground band to the L-band – exactly the number seen in our experiment between 100 and 500 cm$^{-1}$ (Fig. 2b), which is in full agreement with our model.

The transitions indicated by the red arrows in Fig. 4 can now be associated with the peaks in the optical conductivity displayed in Figs. 2 and 3. The wide bump around 25 cm$^{-1}$ corresponds to excitations within the ground band, the lowest set of energy levels, labeled by blue solid arrows in Fig. 4. Although similar excitations are possible within the T and L-bands, we could not unambiguously identify them in our measured spectra due to weak intensity at low temperatures (Boltzmann-distribution governed population). As the temperature rises, the position of 25 cm$^{-1}$ peak slightly shifts to higher energies, by about four wavenumbers when heated from 5 K to 80 K as displayed in Fig. 5. The band vanishes for higher temperatures when all levels are equally populated. The slight temperature shift can be explained by the fact that the potential relief becomes smoother and the tunneling exponent increases. We were not able to identify transitions to higher energy bands because their intensity is too low to be distinguished experimentally.

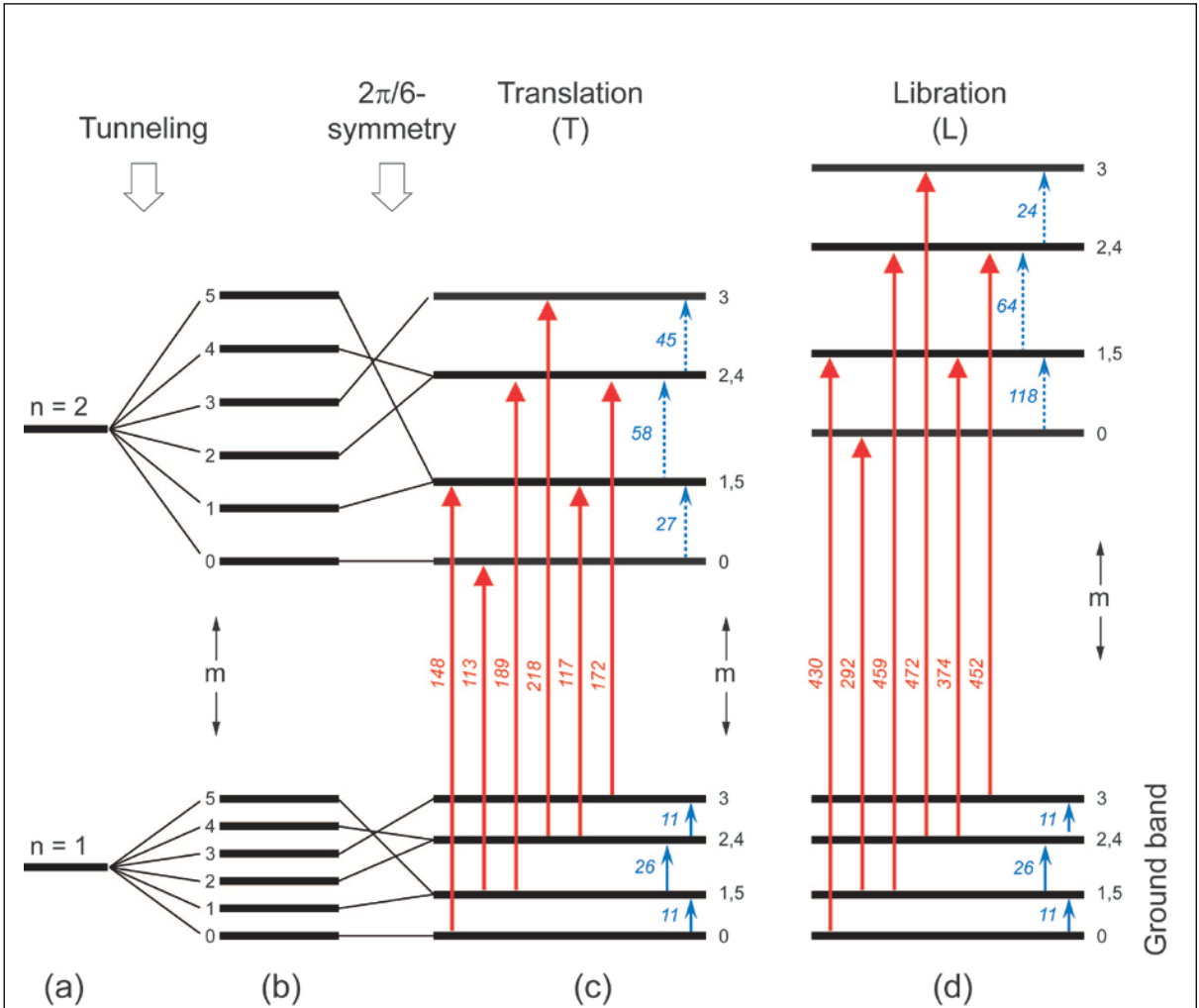

**Figure 4.** Scheme of vibronic energy levels of a type-I water molecule in a nanocavity of the beryl crystal-lattice. The lowest energy bands are shown. **(a)** Ground state (n = 1) and first excited in-plane vibrational state (n = 2) for small vibrations around a preferred direction. **(b)** Due to tunneling within the six-well potential, the states split according to their angular quantum number *m*. Two couples are degenerate: *m* = 1 and *m* = -1 = 5; and also *m* = 2 and *m* = -2 = 4 due to the six-fold symmetry. **(c, d)** The long red arrows indicate inter-band transitions, while the short blue arrows correspond to optical transitions within one band. The selection rule $\Delta m = \pm 1$ allows only certain transitions. The dotted arrows correspond to transitions which are not seen in our experiment. The corresponding frequencies are given in units of cm$^{-1}$, where 100 cm$^{-1}$ corresponds to 12 meV or 3 THz.

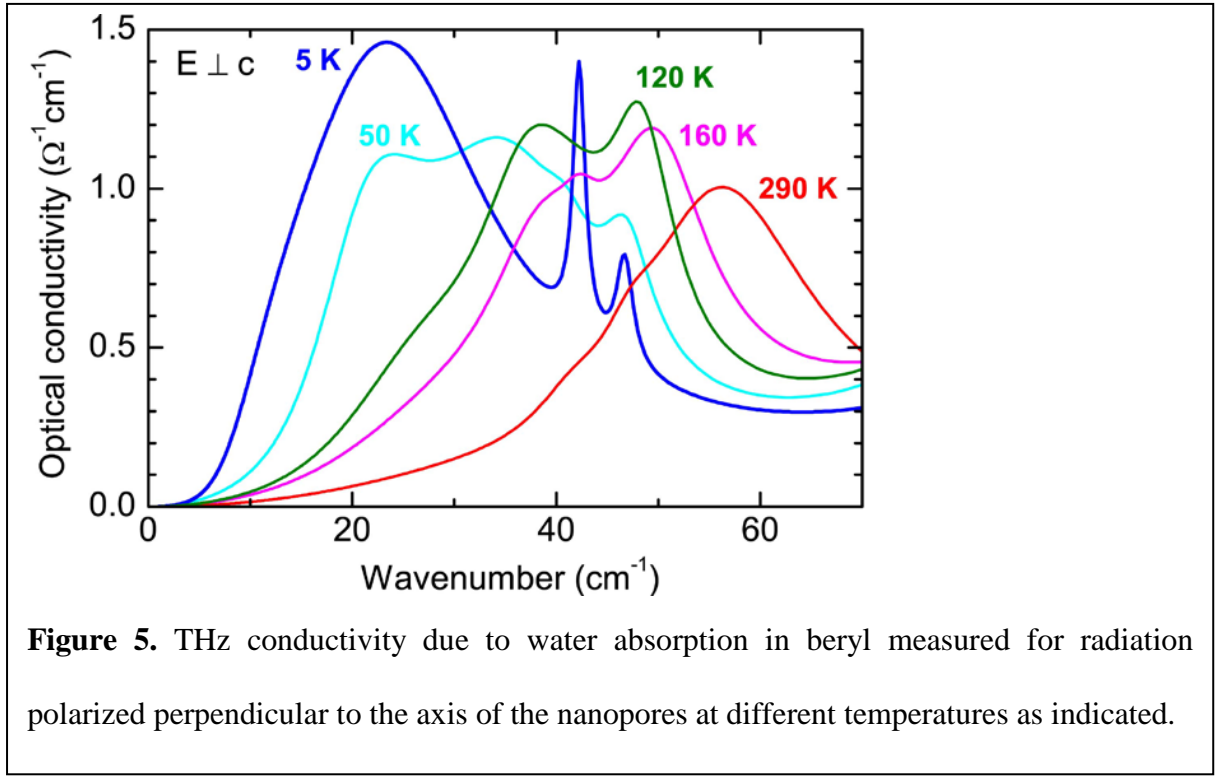

**Figure 5.** THz conductivity due to water absorption in beryl measured for radiation polarized perpendicular to the axis of the nanopores at different temperatures as indicated.

In addition to the broad band at 25 cm$^{-1}$, there are two rather narrow absorption lines observed for $E \perp c$ between 40 and 50 cm$^{-1}$ all the way up to $T = 200$ K, as seen in Fig. 5. They do not fall into a pattern of type-I water vibrations as described above. Their nature can be assigned to the response of type-II water molecules, whose stronger coupling to the cations (compared to type-I molecules that are H-bonded to the cage walls) leads to significantly different characteristics of the resonances, namely smaller damping and spectral weight. Another explanation is based on the observation of similar resonance absorptions around 50 – 60 cm$^{-1}$ in liquid water and ice;[18-21] they might be connected to the bending of the H-bonds, but this is still under debate.[22-25] At low temperatures the absorption lines are very well developed in beryl because the water molecules are confined to a clearly defined and highly symmetric crystalline environment. Our experiments evidence that they do not involve large $H_2O$ molecular complexes, as it is suggested for liquid water or in ice[20-25], and are of extreme local character, i.e. just a single $H_2O$ molecule is involved, in agreement with simulations[22,26] and experiments on aqueous solutions.[27-30] The absorption exhibits a double-peak and is highly sensitive to polarization, which is not seen for the polarization $E \parallel c$. Weak signs of both resonances are detected even at room temperature, as seen in Fig. 5, together with a broad peak located at 56 - 57 $cm^{-1}$.

We have shown that already a weak interaction of a lone water molecule with the walls of a nanosized crystalline cage results in the emergence of a rich set of highly anisotropic molecular vibrational states. By analogy with translational and librational bands in liquid water or ice, corresponding absorption bands can be explained as being due to translational and librational movements of the $H_2O$ molecule which is hydrogen bonded to the cage walls. The six-fold symmetry of the cage, however, causes the T and L bands in beryl to split into a fine structure due to tunnelling within the six-well potential. We believe that the present results will help to analyze more complicated systems with confined water molecules like $H_2O$ chains in carbon nanotubes, molecular clusters in, for example, zeolites, clays, silica gels and other natural or synthetic frameworks and for interfacial water in biological systems.

ACKNOWLEDGMENT

Authors acknowledge fruitful discussions with B.Gompf, C. Holm, K.Lassmann, E. Roduner, A.Simon, L.S.Yaguzhinskii. We thank Dan Wu, N.Aksenov and G.Chanda for their help with the infrared measurements, G.Untereiner, C.Hoch for samples preparation and characterization, G. Siegle and E. Brücher for expert experimental assistance, L. Sebeke and W. Strohmaier for providing the drawings. The research was supported by the RAS Program for fundamental research "Problems of Radiophysics".

SUPPORTING INFORMATION

The sample preparation, the experimental setup used in this study, the analysis of the optical data, the procedure to fit the spectra, and the assignment of the modes for the different polarizations. This information is available free of charge via the Internet at http://pubs.acs.org/.